\documentclass[twocolumn,showpacs,prb]{revtex4-1}%
\usepackage{amsfonts}
\usepackage{amsmath}
\usepackage{amssymb}
\usepackage{subfigure}
\usepackage{graphicx}
\usepackage{txfonts}%
\providecommand{\U}[1]{\protect\rule{.1in}{.1in}}

\begin{document}

\title{Velocity-determined anisotropic behaviors of RKKY interaction in 8-\textit{Pmmn} borophene}
\author{Shu-Hui Zhang$^{1}$}
\email{shuhuizhang@mail.buct.edu.cn}
\author{Ding-Fu Shao$^{2}$}
\author{Wen Yang$^{3}$}
\email{wenyang@csrc.ac.cn}
\affiliation{$^{1}$College of Mathematics and Physics, Beijing University of Chemical Technology, Beijing,
100029, China}
\affiliation{$^{2}$Department of Physics and Astronomy and Nebraska Center for Materials and Nanoscience, University of Nebraska, Lincoln, Nebraska 68588-0299, USA}
\affiliation{$^{3}$Beijing Computational Science Research Center, Beijing 100193, China}

\begin{abstract}
As a new two-dimensional Dirac material, 8-\textit{Pmmn} borophene hosts novel
anisotropic and tilted massless Dirac fermions (MDFs) and has attracted
increasing interest. However, the potential application of 8-\textit{Pmmn}
borophene in spin fields has not been explored. Here, we study the long-range
RKKY interaction mediated by anisotropic and tilted MDFs in magnetically-doped
8-\textit{Pmmn} borophene. To this aim, we carefully analyze the unique
real-space propagation of anisotropic and tilted MDFs with noncolinear momenta
and group velocities. As a result, we analytically demonstrate the anisotropic
behaviors of long-range RKKY interaction, which have no dependence on the Fermi level but are velocity-determined, i.e.,
the anisotropy degrees of oscillation period and envelop amplitude are determined by the anisotropic and tilted velocities. The
velocity-determined RKKY interaction favors to fully determine the characteristic velocities of anisotropic and tilted MDFs through its measurement, and has high tunability by engineering velocities shedding light on the application of
8-\textit{Pmmn} borophene in spin fields.

\end{abstract}
\maketitle


\section{Introduction}

The charge and spin are two intrinsic ingredients of the electron, and the
success of electronics based on the charge transport lures people to develop
the spin-related applications \cite{RevModPhys.76.323,RevModPhys.80.1517}.
However, a lot of solid-state materials are nonmagnetic and/or do not show its
spin properties, then hinder their applications in spin fields. One way to
surmount the obstacle is through the doping of nonmagnetic materials with
magnetic impurity atoms \cite{PowerCrystal2013}. The itinerant carriers of
host materials can help to couple the magnetic impurities indirectly, i.e.,
the Rudermann-Kittel-Kasuya-Yosida (RKKY) interaction
\cite{RudermanPR1954,KasuyaPTP1956,YosidaPR1957}. The RKKY interaction is an
important mechanism underlying rich magnetic phases in diluted magnetic
systems \cite{JungwirthRMP2006} and giant magneto-resistance devices
\cite{RevModPhys.80.1517}, and has potential applications in spintronics
\cite{WolfScience2001,RevModPhys.76.323,MacDonaldNatMater2005}, and\ scalable
quantum computation \cite{LossPRA1998,TrifunovicPRX2012}, as the
RKKY\ interaction enables long-range coupling of distant
spins\cite{PiermarocchiPRL2002,CraigScience2004,RikitakePRB2005,FriesenPRL2007,SrinivasaPRL2015}%
.

The importance of RKKY interaction makes its research to closely accompany the
advent and development of new materials.\ In recent years, the
RKKY\ interaction has been widely studied in various materials such as
graphene
\cite{SaremiPRB2007,BreyPRL2007,HwangPRL2008,SherafatiPRB2011a,KoganPRB2011},
monolayer transition-metal dichalcogenides
\cite{ParhizgarPRB2013b,HatamiPRB2014,MastrogiuseppePRB2014,PhysRevB.93.161404,PhysRevB.94.245429,PhysRevB.99.035107}
topological insulators
\cite{LiuPRL2009,ZhuPRL2011,AbaninPRL2011,PhysRevB.96.081405,PhysRevB.97.125432,1812.02281}%
, silicene\cite{ZarePRB2016,Duan2018},
phosphorene\cite{1367-2630-19-10-103010,PhysRevB.98.205401,PhysRevB.97.235424}, layered structures \cite{Masrour2016,Jabar2017,Jabar2018,Jabar2019}
and Dirac and Weyl semimetals
\cite{HosseiniPRB2015,ChangPRB2015,PhysRevB.93.094433}. Among these studies,
the RKKY interaction mediated by massless Dirac fermions (MDFs) has attracted
a lot of interest, which exhibits many new behaviors and brings about
potential opportunities of nonmagnetic Dirac materials for spin applications.
However, previous studies are mainly limited to the Dirac materials with
isotropic MDFs, this leaves the influence of novel MDFs on the RKKY
interaction unexplored.

\begin{figure}[ptbh]
\includegraphics[width=1.0\columnwidth,clip]{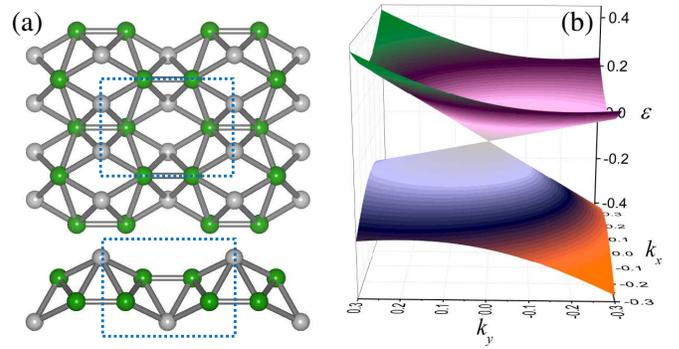}\caption{(a) Crystal structure of 8-Pmmn borophene from top view (upper panel) and side view (bottom panel). The ridge boron atom (green) and the inner
boron atom (gray) are nonequivalent. In each unit cell, there are eight atoms delimited by the dashed blue rectangle. (b) Anisotropic and
tilted Dirac cone in 8-\textit{Pmmn} borophene. }%
\label{ES}%
\end{figure}

The 8-\textit{Pmmn} borophene with the crystal structure of Fig. \ref{ES}(a) is one kind of two-dimensional Dirac material,
in which\ the novel MDFs are anisotropic and tilted as shown by Fig.
\ref{ES}(b) in contrast to those isotropic ones in the well-known graphene.
Recently, since its seminal prediction \cite{PhysRevLett.112.085502}, wide
theoretical efforts have been paid to calculate the unprecedented electronic
properties
\cite{PhysRevB.93.241405,PhysRevB.94.165403,PhysRevB.97.125424,PhysRevB.98.115413}
and to construct the effective low-energy continuum Hamiltonian
\cite{PhysRevB.94.165403,PhysRevB.97.125424}\ which has been used to study the
plasmon dispersion and screening properties \cite{PhysRevB.96.035410}, the
optical conductivity \cite{PhysRevB.96.155418}, Weiss oscillations
\cite{PhysRevB.96.235405}, and Metal-insulator transition induced by strong electromagnetic radiation \cite{PhysRevB.99.035415}. The rapid experimental advances of various
borophene monolayers\ \cite{MannixS1513,FengNC2016,PhysRevLett.118.096401}
further boost wider research interest on 8-\textit{Pmmn} borophene. The anisotropic exchange coupling and the magnetic anisotropy are very crucial to spin applications\cite{RevModPhys.76.323,JungwirthRMP2006,Shi2018,ShiJiang2018,ZouGuo2018}, so our interest is to examine the anisotropic features of RKKY interaction due to the novel MDFs.. Here, we
study the RKKY interaction mediated by the anisotropic and tilted MDFs by
using the effective model of 8-\textit{Pmmn} borophene. In the long range, we
analytically derive the Green's function (GF) of anisotropic and tilted MDFs
to show the unique real-space propagation properties, and then their mediation
to RKKY interaction whose oscillation period and envelop amplitude are both
anisotropic and determined by the velocities, i.e., velocity-determined RKKY
interaction. As a result, our theoretical study show the potential of
velocity-determined RKKY interaction in the characterization of characteristic velocities of anisotropic tilted MDFs and the application of 8-\textit{Pmmn} borophene for spin fields.

This paper is organized as follows. In Sec. II, we start from the intrinsic
electronic properties of 8-\textit{Pmmn} borophene, present the general
expression of RKKY interaction, and reveal the analytical behaviors of the
real-space GF including the symmetry properties, the classical trajectories
and the explicit analytical expression. Then in Sec. III, we perform exact
numerical calculations to show the typical features of velocity-determined
RKKY interaction. Most importantly, the analytical expression for RKKY is
derived and is used to understand the velocity-determined RKKY interaction.
Finally, we summarize this study in Sec. IV.

\section{Theoretical formalism}

In 8-\textit{Pmmn} borophene, the effective Hamiltonian of anisotropic and
tilted MDFs near one Dirac cone is
\cite{PhysRevB.94.165403,PhysRevB.96.235405}%

\begin{equation}
\hat{H}_{0}=(v_{x}\sigma_{x}\hat{p}_{x}+v_{y}\sigma_{y}\hat{p}_{y}+v_{t}%
\sigma_{0}\hat{p}_{y}),
\end{equation}
where $\hat{p}_{x,y}$ are the momentum operators, $\sigma_{x,y}$ are
$2\times2$ Pauli matrices, and $\sigma_{0}$ is the $2\times2$ identity matrix.
The anisotropic velocities are $v_{x}=0.86v_{F}$, $v_{y}=0.69v_{F}$, and
$v_{t}=0.32v_{F}$ with $v_{F}=10^{6}$ m/s. In this study, we set $\hbar
=v_{F}\equiv1$ to favor our dimensionless derivation and calculations since
they can define the length unit $l_{0}$\ and the energy unit $\varepsilon_{0}$
through $\hbar v_{F}=l_{0}\varepsilon_{0}$, e.g., $\varepsilon_{0}=0.66$ eV
when $l_{0}=1$ nm. To solve the Sch\"{o}dinger equation $\hat{H}_{0}%
\psi=\varepsilon\psi$, the energy dispersion [see Fig. \ref{ES}(b)] and the
corresponding wave functions\ are, respectively,%

\begin{equation}
\varepsilon_{\lambda,\mathbf{k}}=v_{t}k_{y}+\lambda\sqrt{v_{x}^{2}k_{x}%
^{2}+v_{y}^{2}k_{y}^{2}}, \label{ED}%
\end{equation}
and%

\begin{equation}
\psi_{\lambda,\mathbf{k}}(\mathbf{r})=e^{i\mathbf{k}\cdot\mathbf{r}%
}|u_{\lambda}(k_{x},k_{y})\rangle
\end{equation}
with
\[
|u_{\lambda}(k_{x},k_{y})\rangle\equiv\frac{1}{\sqrt{2}}%
\begin{bmatrix}
1\\
\frac{\left(  v_{x}k_{x}+iv_{y}k_{y}\right)  }{\varepsilon_{\lambda
,\mathbf{k}}-v_{t}k_{y}}%
\end{bmatrix}
.
\]
Here, $\lambda=\pm$ denotes the conduction (valence) band, $\mathbf{k=}%
(k_{x},k_{y})$ and $\mathbf{r=}(x,y)$ are, respectively, the momentum and
position vectors, and $|u_{\lambda}(k_{x},k_{y})\rangle$ is the spinor part of
the wave function.

\begin{figure}[ptbh]
\includegraphics[width=\columnwidth,clip]{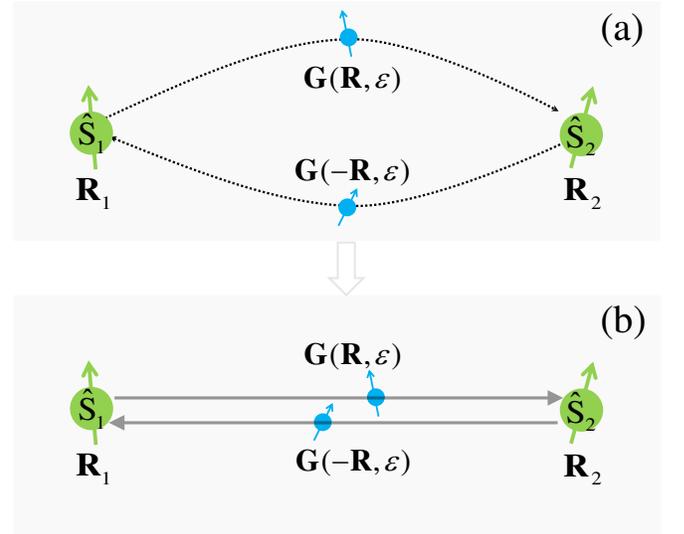}\caption{Schematic diagram
of RKKY interaction of two localized spins $\hat{\mathbf{S}}_{1,2}$\ at
$\mathbf{R}_{1,2}$ mediated by the propagation Green's functions
$\mathbf{G}(\mathbf{R,}\varepsilon)$ and $\mathbf{G}(-\mathbf{R},\varepsilon)$
of itinerant carriers between them. Generally, all quantum states below the
Fermi level $\varepsilon_{F}$ (see Eq. \ref{RF}) contribute to the Green's
functions as shown by the curved dotted lines in (a), and then the RKKY
interaction. However, at large distances, the Green's functions are dominated
by the states with group velocities parallel to $\mathbf{R}$ (i.e., the
classical trajectories as shown by the straight solid lines in (b)), and then
RKKY interaction is determined by the limited momentum states on the Fermi
surface (see Eq. \ref{ARF}).}%
\label{RKKY}%
\end{figure}

\subsection{Expression for the RKKY interaction}

In the uniform 8-\textit{Pmmn} borophene, as shown by Fig. \ref{RKKY}(a),\ for
two local spins $\hat{\mathbf{S}}_{1,2}$ at $\mathbf{R}_{1,2}$ coupled to the
carrier spin density $\hat{\mathbf{s}}(\mathbf{x})=\hat{\mathbf{s}}\delta
(\hat{\mathbf{r}}-\mathbf{x})$ via the s-d exchange interaction $-J_{0}%
\hat{\mathbf{S}}_{1}\cdot\hat{\mathbf{s}}(\mathbf{R}_{1})-J_{0}\hat
{\mathbf{S}}_{2}\cdot\hat{\mathbf{s}}(\mathbf{R}_{2})$, the carrier-mediated
RKKY interaction assumes the isotropic Heisenberg form due to the absence of
spin-orbit coupling, i.e., $\hat{H}_{\text{RKKY}}=J_{jj^{\prime}}%
\mathbf{\hat{S}}_{1}\cdot\mathbf{\hat{S}}_{2}$, where the range function
\cite{SherafatiPRB2011a,KlinovajaPRB2013,zhang2017}%

\begin{equation}
J_{jj^{\prime}}(\varepsilon_{F},\mathbf{R})=-\frac{J_{0}^{2}}{2\pi
}\operatorname{Im}\int_{-\infty}^{\varepsilon_{F}}\digamma_{jj^{\prime}%
}(\varepsilon,\mathbf{R})d\varepsilon\label{RF}%
\end{equation}
is determined by the intrinsic real-space GF of carriers, i.e.,
\begin{equation}
\digamma_{jj^{\prime}}(\varepsilon,\mathbf{R})=G_{jj^{\prime}}(\mathbf{R}%
,\varepsilon)G_{j^{\prime}j}(-\mathbf{R},\varepsilon). \label{MG2}%
\end{equation}
Due to the integral forms of Eq. \ref{RF} over the energy and of Eq.
\ref{2DGF}\ over the momentum (see the subsequent subsection), all quantum
states propagating between $\mathbf{R}_{1}$ and $\mathbf{R}_{2}$ below the
Fermi level $\varepsilon_{F}$ contribute to the RKKY interaction as shown by
the curved dotted line in Fig. \ref{RKKY}(a). In addition, $j,j^{\prime}%
\in\{1,2\}$ label the orbitals used in the basis for the Hamiltonian of
8-\textit{Pmmn} borophene. Until now, the respective contribution of different
orbitals to form the Dirac states of 8-\textit{Pmmn} borophene is still in
debate \cite{PhysRevB.94.165403} and the interaction properties between
magnetic impurity and the borons' orbitals call for further density functional
calculations, so we focus on the long-range behaviors of RKKY interaction in
our model study.

\subsection{Real-space GF}

The real-space GF is the key to understand the spatial propagation of carries
and their mediation to the RKKY interaction, which is defined by%

\begin{equation}
\mathbf{G}(\mathbf{R},\varepsilon)\equiv\langle\mathbf{R}_{2}|(\varepsilon
+i0^{+}-\hat{H}_{0})^{-1}|\mathbf{R}_{1}\rangle.
\end{equation}
Here, $\mathbf{R=R}_{2}-\mathbf{R}_{1}\ $accounts for the translational
invariance in uniform system and is the connecting vector for two magnetic
impurities, and $\mathbf{G}$ is one $2\times2$ matrix expressed in the orbital
basis instead of the spin basis originating from the spin-degenerate
Hamiltonian of 8-\textit{Pmmn} borophene in the orbital
basis\cite{PhysRevB.94.165403}.\ The real-space GF can be obtained from the
Fourier transformation%

\begin{equation}
\mathbf{G}(\mathbf{R},\varepsilon)=\frac{1}{4\pi^{2}}%
{\displaystyle\iint}
d^{2}\mathbf{k}e^{i\mathbf{k}\cdot\mathbf{R}}\mathbf{G}(\mathbf{k}%
,\varepsilon), \label{2DGF}%
\end{equation}
of the momentum-space GF%

\begin{equation}
\mathbf{G}(\mathbf{k},\varepsilon)\equiv\lbrack z-H_{0}(\mathbf{k}%
)]^{-1}=-\frac{2(\epsilon-\gamma_{2}k_{y})}{v_{x}}\frac{|u_{\lambda}%
(k_{x},k_{y})\rangle\langle u_{\lambda}(k_{x},k_{y})|}{(k_{x}-k_{x,+}%
)(k_{x}-k_{x,-})}.
\end{equation}
with%

\begin{equation}
k_{x,\pm}(k_{y})=\pm\sqrt{(\epsilon-\gamma_{2}k_{y})^{2}-\gamma_{1}^{2}%
k_{y}^{2}}.
\end{equation}
Here, $\gamma_{1}=v_{y}/v_{x}$, $\gamma_{2}=v_{t}/v_{x}$, $\epsilon=z/v_{x}$
with $z=\varepsilon+i0^{+}$, and $\lambda=\mathrm{sgn}(\epsilon)$ conforms to
the energy dispersion. In the definition of momentum-space GF, we have used
the form of $\hat{H}_{0}$ in the momentum space, i.e., $H_{0}(\mathbf{k}%
)=\langle\mathbf{k}|\hat{H}_{0}|\mathbf{k}\rangle$. To perform the contour
integral over $k_{x}$ straightforwardly:%

\begin{equation}
\mathbf{G}(\mathbf{R},\varepsilon)=\frac{\mathrm{sgn}(x)}{2\pi}\int
dk_{y}e^{i\phi_{s}}\frac{\epsilon-\gamma_{2}k_{y}}{iv_{x}k_{x,s}}|u_{\lambda
}(k_{x},k_{y})\rangle\langle u_{\lambda}(k_{x},k_{y})|, \label{1DGF}%
\end{equation}
where $s=\mathrm{sgn}(x)\mathrm{sgn}(\epsilon-\gamma_{2}k_{y})$, and $\phi
_{s}(k_{y},\mathbf{R})=k_{x,s}x+k_{y}y$ is the propagation phase factor which
has an energy dependence but is omitted for conciseness. And noting here, we
always adopt $x=0^{+}$ to replace $x=0$ when we consider $\mathbf{R}$ along
the $y$-axis. Later, we firstly discuss the symmetry properties of GF, then
derive the dominant contributing states (namely, the classical
trajectories)\ to real-space GF and their features, which help us to derive
the analytical expression of real-space GF.

\subsubsection{Symmetry properties of GF}

From the Eq. \ref{1DGF}, the symmetry properties of GF can be derived, which
determines the RKKY properties. On one hand, we have $G_{11}(\mathbf{R}%
,\varepsilon)=G_{22}(\mathbf{R},\varepsilon)$, this is identical to the case
of graphene with isotropic MDFs \cite{SherafatiPRB2011a}. However, there is no
symmetrical relation between $G_{12}(\mathbf{R},\varepsilon)$ and
$G_{21}(\mathbf{R},\varepsilon)$ in contrast to $G_{12}(\mathbf{R}%
,\varepsilon)=G_{21}(\mathbf{R},\varepsilon)$ in graphene \cite{zhang2017}. On
the other hand, one can arrive at
\begin{equation}
\mathbf{G}(\mathbf{R},-\varepsilon)=-(\mathbf{G}^{\ast}(\mathbf{R}%
,\varepsilon))^{\text{T}}. \label{EHS}%
\end{equation}
Here, T denotes the transpose operation of a matrix. Eq. \ref{EHS} is also
applicable to isotropic MDFs in graphene as expected from the electron-hole
symmetry \cite{zhang2017}, so its origin\ can be seen as a generalized
electron-hole symmetry for anisotropic and tilted MDFs. The symmetry
properties is beneficial to the discussions of\ the RKKY interaction and
reduces the numerical calculations as shown in the next section, e.g., our
focus can be mainly on the electron-doped case by using Eq. \ref{EHS}.

\subsubsection{Classical trajectory: traveling nature and group velocity}

The integral form of Eq. \ref{1DGF} over $k_{y}$ shows that in general all
momentum states should contribute together to the real-space GF. For
convenience, we\ can separate these momentum states into two classes:
traveling states and evanescent states. For the conductance (valence)\ band
with $\lambda=+$ ($\lambda=-$), we have traveling states with $\gamma_{2}%
k_{y}<\epsilon$ ($\gamma_{2}k_{y}>\epsilon$) and evanescent states
with\ $\gamma_{2}k_{y}>\epsilon$ ($\gamma_{2}k_{y}<\epsilon$). To identify the
dominant contributing states to GF, we use the saddle point approximation.
According to the saddle point approximation
\cite{RothPR1966,LiuPRB2012a,PowerPRB2011,LounisPRB2011}, the dominant
contribution to Eq. \ref{1DGF} is the classical trajectory from $\mathbf{R}%
_{1}$ to $\mathbf{R}_{2}$, which is determined by the first order derivative
of propagation phase factor $\phi_{s}$:
\begin{equation}
\frac{\partial\phi_{s}}{\partial k_{y}}=0. \label{SPA}%
\end{equation}
Eq. \ref{SPA} leads to the identity:%

\begin{equation}
\frac{\gamma_{2}(\epsilon-k_{y,s}^{c}\gamma_{2})+\gamma_{1}^{2}k_{y,s}^{c}%
}{k_{x,s}^{c}}=\tan\theta, \label{CLA}%
\end{equation}
with $\tan\theta=y/x$ and $\theta$ being the azimuthal angle of $\mathbf{R}$
elative to the $x$ axis. Eq. \ref{CLA} can be rewritten as a quadratic
equation with one unknown:%

\begin{equation}
(\allowbreak\gamma_{1}^{2}-\gamma_{2}^{2})(k_{y,s}^{c})^{2}+\allowbreak
2\epsilon\gamma_{2}k_{y,s}^{c}+\frac{\epsilon^{2}\left(  \gamma_{2}^{2}%
-\tan^{2}\theta\right)  \allowbreak}{\gamma_{1}^{2}-\gamma_{2}^{2}+\tan
^{2}\theta}=0.
\end{equation}
Once $k_{y,s}^{c}$ are derived, one can obtain $k_{x,s}^{c}=k_{x,s}%
(k_{y,s}^{c})$ for the classical trajectory.\ In other words, the classical
trajectory has the momentum $\mathbf{k}_{s}^{c}=(k_{x,s}^{c},k_{y,s}^{c})$, in
which the components have explicit analytical expressions:%

\begin{subequations}
\label{AKC}%
\begin{align}
k_{x,s}^{c}  &  =\frac{s|\epsilon|\gamma_{1}}{\sqrt{\gamma_{1}^{2}-\gamma
_{2}^{2}+\tan^{2}\theta}},\\
k_{y,s}^{c}  &  =-\frac{1}{\gamma_{1}^{2}-\gamma_{2}^{2}}(\epsilon\gamma
_{2}-\frac{s|\epsilon|\gamma_{1}\tan\theta}{\sqrt{\gamma_{1}^{2}-\gamma
_{2}^{2}+\tan^{2}\theta}}).
\end{align}
So one can arrive at $\epsilon-\gamma_{2}k_{y,s}^{c}=\epsilon f(\gamma
_{1},\gamma_{2},\theta)$ with \ %

\end{subequations}
\begin{equation}
f(\gamma_{1},\gamma_{2},\theta)=\frac{\gamma_{1}^{2}}{\gamma_{1}^{2}%
-\gamma_{2}^{2}}-\frac{s\lambda\gamma_{1}\gamma_{2}}{\gamma_{1}^{2}-\gamma
_{2}^{2}}\frac{\tan\theta}{\sqrt{\gamma_{1}^{2}-\gamma_{2}^{2}+\tan^{2}\theta
}}.
\end{equation}
For the anisotropic and tilted MDFs with $\gamma_{1}>\gamma_{2}$ in
8-\textit{Pmmn} borophene, $f(\gamma_{1},\gamma_{2},\theta)>\gamma_{1}%
/(\gamma_{1}+\gamma_{2})>0$ leads to $\mathrm{sgn}(\epsilon-\gamma_{2}%
k_{y,s}^{c})=\lambda$ or $s=\lambda\mathrm{sgn(}x$) for the classical
trajectory. Therefore, the classical trajectory is always of traveling nature.

\begin{figure}[ptbh]
\includegraphics[width=0.9\columnwidth,clip]{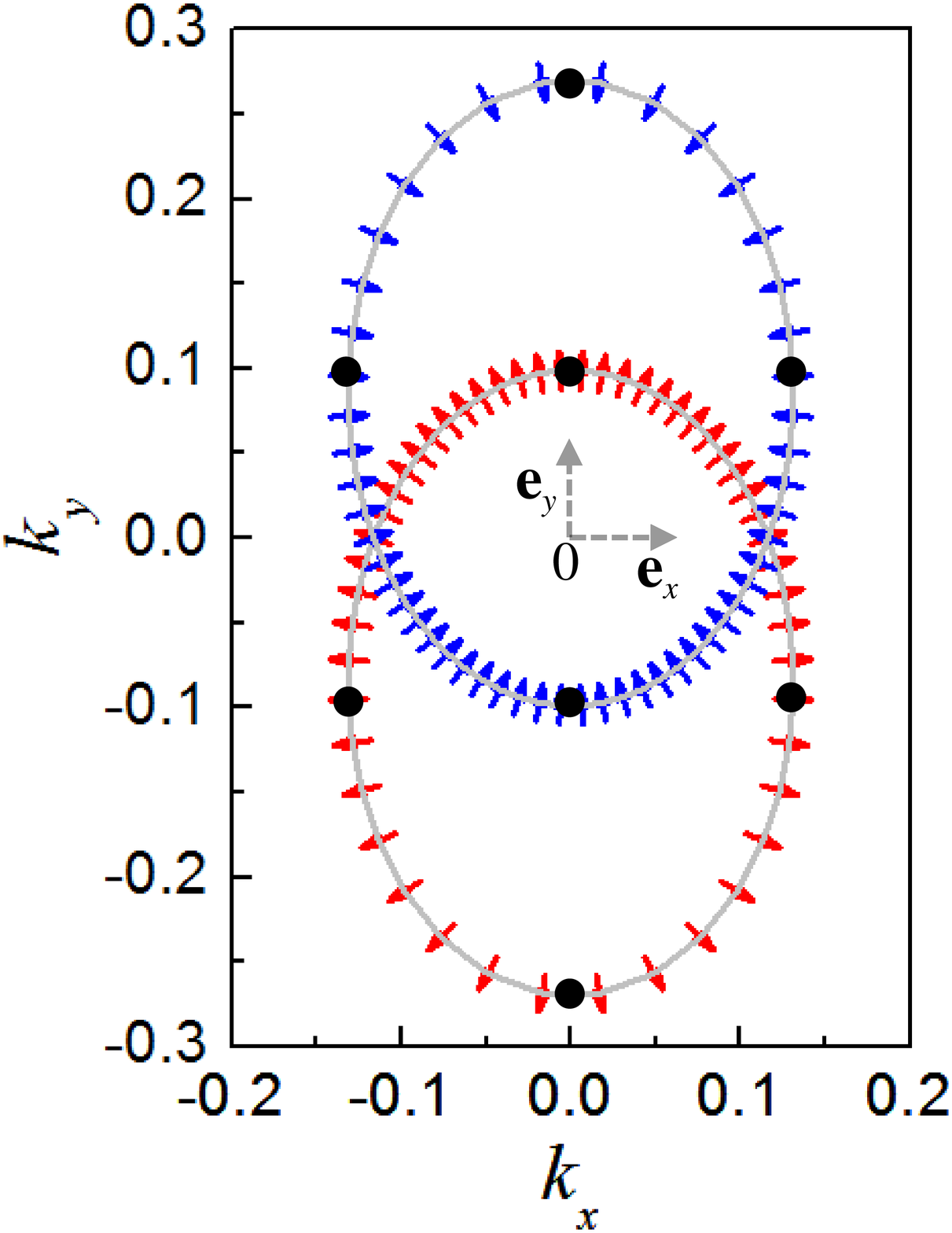}\caption{Texture of
electron (red)\ and hole (blue) group velocities on the Fermi surfaces with
$\varepsilon_{+,\mathbf{k}}=0.1$ and $\varepsilon_{-,\mathbf{k}}=-0.1$. The
Fermi surfaces are shifted ellipses with noncolinear momenta and group
velocities, and the group velocities (black dots) on the vertices of semimajor
(semiminor) axis are parallel to ${\mathbf{e}}_{y}$ (${\mathbf{e}}_{x}$) of
the Cartesian coordinate system $x-y$.}%
\label{vcontour}%
\end{figure}

Turning to the group velocity $\mathbf{v}_{\lambda}(\varepsilon,\mathbf{k}%
)=(v_{\lambda,x},v_{\lambda,y})$ of anisotropic tilted MDFs, which has the components%

\begin{subequations}
\label{GVEC}%
\begin{align}
v_{\lambda,x}  &  \equiv\frac{\partial\varepsilon_{\lambda,\mathbf{k}}%
}{\partial k_{x}}=\frac{\lambda v_{x}k_{x}}{\sqrt{k_{x}^{2}+\gamma_{1}%
^{2}k_{y}^{2}}},\\
v_{\lambda,y}  &  \equiv\frac{\partial\varepsilon_{\lambda,\mathbf{k}}%
}{\partial k_{y}}=v_{t}+\frac{\lambda v_{x}\gamma_{1}^{2}k_{y}}{\sqrt
{k_{x}^{2}+\gamma_{1}^{2}k_{y}^{2}}}.
\end{align}
Eq. \ref{GVEC} shows that the electron and hole group velocities are related
to each other through $\mathbf{v}_{+}(\varepsilon,\mathbf{k})=\mathbf{v}%
_{-}(-\varepsilon,-\mathbf{k})$. In Fig. \ref{vcontour}, we show the texture
of electron (red color)\ and hole (blue color) group velocities on the Fermi
surfaces with $\varepsilon_{+,\mathbf{k}}=0.1$ and $\varepsilon_{-,\mathbf{k}%
}=-0.1$. Due to the anisotropy and tilt, the Fermi surface is a shifted
ellipse with its semimajor (semiminor) axis parallel to ${\mathbf{e}}_{y}$
(${\mathbf{e}}_{x}$) of the Cartesian coordinate system $x-y$, and has
noncolinear momenta and group velocities \cite{PhysRevB.97.235440}. The
velocity textures for electrons and holes shown by Fig. \ref{vcontour} have
the inversion symmetry in the momentum space consistent with Eq. \ref{GVEC}.

On the classical trajectory with the momentum $\mathbf{k}_{s}^{c}=(k_{x,s}%
^{c},k_{y,s}^{c})$, one can obtain%

\end{subequations}
\begin{equation}
\frac{v_{\lambda,y}^{c}}{v_{\lambda,x}^{c}}=\tan\theta.
\end{equation}
As a result, $\mathbf{v}_{\lambda}^{c}(\varepsilon,\mathbf{k}_{s}%
^{c})\mathbf{\parallel R}$ for the classical trajectory, this implies
$\mathbf{G}(\mathbf{R},\varepsilon)$\ are mainly contributed by the states
with the group velocities parallel to the connecting vector $\mathbf{R}$. For
convenience, we introduce $\mathbf{v}_{\lambda}^{c}(\varepsilon,\mathbf{R}%
)\equiv\mathbf{v}_{\lambda}^{c}(\varepsilon,\mathbf{k}_{s}^{c})$.
$\mathbf{v}_{\lambda}^{c}(\varepsilon,\mathbf{R})\mathbf{\parallel R}$ for the
classical trajectory provides one principle to identify the dominant momentum
states contributing to real-space GF $\mathbf{G}(\mathbf{R},\varepsilon)$,
i.e.,\ through comparing the direction of the group velocity of each state on
the Fermi surface and that of $\mathbf{R}$\textbf{. }As an application of this
skill, we give an example to consider $\mathbf{G}(\mathbf{R},\varepsilon)$ for
electrons without loss of generality. According to the velocity texture shown
in Fig. \ref{vcontour}, the dominant contributing momentum states on the
electron Fermi surface lie on the right (left) vertex of semiminor axis for
$\mathbf{G}(\mathbf{R},\varepsilon)$ along the positive (negative) $x$-axis
which is consistent with Eq. \ref{AKC} with $\theta=0$ and $\pi$:%

\begin{subequations}
\label{XXTK}%
\begin{align}
k_{x,\pm}^{c}  &  =\pm\frac{\epsilon\gamma_{1}}{\sqrt{\gamma_{1}^{2}%
-\gamma_{2}^{2}}},\\
k_{y,\pm}^{c}  &  =-\frac{\epsilon\gamma_{2}}{\gamma_{1}^{2}-\gamma_{2}^{2}}.
\end{align}
Noting here $k_{y,\pm}^{c}$ adopt one value and equal to the shifted momentum
of the center of the elliptical Fermi surface away from the coordinate
origin\cite{PhysRevB.97.235440}, which helps to determine the momentum
position of semiminor axis in the $y$ direction. Similarly, the dominant
contributing momentum states on the electron Fermi surface lie on the upper
(bottom) vertex of semmajor axis for $\mathbf{G}(\mathbf{R},\varepsilon)$
along the positive (negative) $y$-axis, which is consistent with Eq. \ref{AKC}
with $\theta=\pm\pi/2$:%

\end{subequations}
\begin{subequations}
\label{YYTK}%
\begin{align}
k_{x,\pm}^{c}  &  =0,\\
k_{y,\pm}^{c}  &  =\frac{\epsilon}{\gamma_{2}\pm\gamma_{1}}.
\end{align}
Therefore, it is an effective principle to identify the dominant contributing
momentum states to GF by using the velocity texture on the Fermi surface, and
this principle should be very useful for the complex Fermi surfaces without
analytical solutions.

\subsubsection{Analytical GF}

Near the classical trajectory, the Taylor expansion of $\phi_{s}(k_{y})$ is%

\end{subequations}
\begin{equation}
\phi_{s}(k_{y},\mathbf{R})=\phi_{s}^{c}(\mathbf{R})-\alpha_{s}(k_{y}%
-k_{y,s}^{c})^{2}+o(k_{y}),
\end{equation}
where $\phi_{s}^{c}(\mathbf{R})\equiv\phi_{s}^{c}(k_{y,s}^{c},\mathbf{R})$
since the $\mathbf{R}$ determines $k_{y,s}^{c}$ for a fixed Fermi surface,
$o(k_{y})$ is the high-order small quantity as the function of $k_{y}$ and
\begin{equation}
\alpha_{s}=-\frac{1}{2}\frac{\partial^{2}\phi_{s}}{\partial k_{y}^{2}%
}|_{k_{y,s}^{c}}=\frac{x\epsilon^{2}\gamma_{1}^{2}}{2(k_{x,s}^{c})^{3}}.
\label{alphas}%
\end{equation}
In light of the stationary phase approximation, the real-space GF at the large
distance is contributed mainly by the electron propagation along the classical
trajectory. As a result, the real-space GF can be expressed analytically as:%

\begin{subequations}
\label{AGF}%
\begin{align}
\mathbf{G}(\mathbf{R},\varepsilon)  &  \approx\frac{\text{sgn(}x\text{)}%
e^{i\phi_{s}^{c}}}{2\pi}\frac{\epsilon-\gamma_{2}k_{y,s}^{c}}{iv_{x}%
k_{x,s}^{c}}\sqrt{\frac{\pi}{i\alpha_{s}}}|u_{\lambda}(k_{x,s}^{c},k_{y,s}%
^{c})\rangle\langle u_{\lambda}(k_{x,s}^{c},k_{y,s}^{c})|,\\
&  =\frac{e^{i\phi_{s}^{c}-\lambda\pi/4}}{2\pi}\frac{\epsilon-\gamma
_{2}k_{y,s}^{c}}{i\lambda v_{x}|k_{x,s}^{c}|}\sqrt{\frac{\pi}{|\alpha_{s}|}%
}|u_{\lambda}(k_{x,s}^{c},k_{y,s}^{c})\rangle\langle u_{\lambda}(k_{x,s}%
^{c},k_{y,s}^{c})|,
\end{align}
Here, the Gaussian formula $\int_{-\infty}^{\infty}e^{-\alpha x^{2}}%
dx=\sqrt{\pi/\alpha}$ has been used. And $\alpha$ is a complex variable with
the angle $\arg(\alpha)\in(-\pi,\pi]$, so $\sqrt{\alpha}=\sqrt{\left\vert
\alpha\right\vert }e^{i\arg(\alpha)/2}$. It is worthy to emphasize that Eq.
\ref{AGF} is consistent with Eq. \ref{EHS} and has a clear symmetry origin,
i.e., the inversion symmetry of velocity textures for electrons and holes as
shown by Fig. \ref{vcontour}.

\section{Results and discussions}

In order to clearly present the numerical and analytical results for the RKKY
interaction in 8-\textit{Pmmn} borophene, we introduce the scaled RKKY range
function ${\mathcal{J}}_{jj^{\prime}}=J_{jj^{\prime}}R^{2}/J_{0}^{2}$ with
$R=\left\vert \mathbf{R}\right\vert $. Fig. \ref{CRKKY}(a) and (b) show the
RKKY range function ${\mathcal{J}}_{11}$\ for the connection vector
$\mathbf{R}$ between two impurities along $x$-axis and $y$-axis, in which the
red dashed lines are plotted by using the results of exact numerical
calculations based on the Eqs. \ref{RF} and \ref{1DGF}.

\subsection{Negligible sublattice and valley effect}

In the unit cell of 8-\textit{Pmmn} borophene, there are 8 boron atoms, among
which two are nonequivalent and maybe regarded as the sublattice degree of
freedom in analogy to graphene\cite{SherafatiPRB2011a,PhysRevB.92.235435}.
Now, the respective contribution of different orbitals of different atoms to
form the Dirac states of 8-\textit{Pmmn} borophene is still in
debate\cite{PhysRevB.94.165403}. The low-energy spectrum of 8-\textit{Pmmn}
borophene is expressed in a proper basis, which has no clear connection with
the nonequivalent atoms in contrast to the case in graphene. So using the
low-energy spectrum of 8-\textit{Pmmn} borophene, we can just discuss the
orbital-dependent RKKY interaction as done in three-dimensional Dirac
semimetals\cite{PhysRevB.93.094433}.

The carrier-mediated RKKY interaction is determined by the GF of
carriers (cf. Eq. \ref{RF}). The 8-\textit{Pmmn} borophene has two
inequivalent Dirac cones which can be included in the calculation of GF by combing two valley-dependent GFs. The combination
should be a sum of valley-dependent GFs weighed by proper phase
factors\cite{Bena2009,SherafatiPRB2011a}.\ The phase factors can be identified
through the low-energy expansion of the lattice model. At present, the lattice
model for 8-\textit{Pmmn} borophene is very tedious\cite{PhysRevB.94.165403},
so the phase factors are not known. Even if two valleys can be included
properly, their interference only cause the short-range oscillation decorated
on the RKKY interaction as a function of impurity distance. Due to the large
momentum spacing between two valleys\cite{PhysRevB.94.165403}, the short-range
oscillation is on the atomic scale.

Here, we consider the RKKY interaction in the doped 8-\textit{Pmmn} borophene,
and focus its long-range oscillation determined by the Fermi wavelength on the
nanometer scale, decay rate determined by the system dimension and the envelop
amplitude without dependence on the oscillation detail. Therefore, the
sublattice effect and valley effect are negligible in our study.

\begin{figure}[ptbh]
\includegraphics[width=\columnwidth,clip]{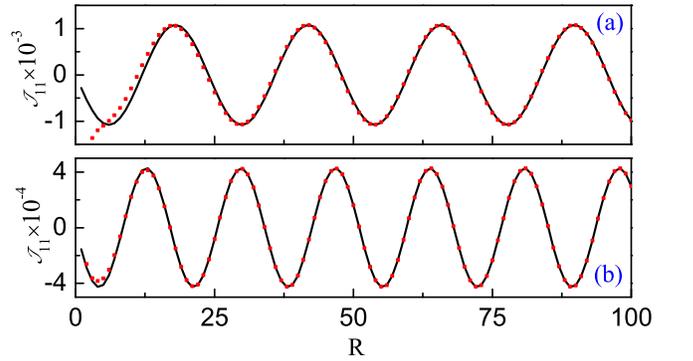}\caption{RKKY range
function ${\mathcal{J}}_{11}$\ for the connection vector between two
impurities along $x$-axis (a)\ and $y$-axis (b). In the plots, we have used
the analytical formula (black solid lines) and numerical calculations (red
dotted lines) for the electron doping $\varepsilon_{F}=0.1$. }%
\label{CRKKY}
\end{figure}

\subsection{Analytical expressions for RKKY interaction}

To understand the unique behaviors of RKKY interaction mediated by anisotropic
and tilted MDFs in 8-\textit{Pmmn} borophene, we derive the analytical formula
for $J_{jj^{\prime}}$. At large distances, the integrand in\ Eq. \ref{RF}\ can
be rewritten as%

\end{subequations}
\begin{equation}
\digamma_{jj^{\prime}}(\varepsilon,\mathbf{R})=\digamma_{jj^{\prime}}%
^{0}(\varepsilon,\mathbf{R})e^{i\phi_{s}(\mathbf{R})+i\phi_{-s}(-\mathbf{R})}%
\end{equation}
where $\digamma_{jj^{\prime}}^{0}(\varepsilon,\mathbf{R})$ is a slowly-varying
function of the energy. Due to the rapid oscillation of the phase factor
$e^{i\phi_{s}(\mathbf{R})+i\phi_{-s}(-\mathbf{R})}$, the integral over energy
in Eq. \ref{RF} is dominated by the contributions near the Fermi energy
$\varepsilon_{F}$, so we can perform the expansion as follows:%

\begin{equation}
\phi_{s}(\mathbf{R})+\phi_{-s}(-\mathbf{R})\approx\phi_{s}^{c}(\mathbf{R}%
)+\phi_{-s}^{c}(-\mathbf{R})+\frac{R}{\bar{v}_{\lambda}^{c}(\varepsilon_{F}%
)}(\varepsilon-\varepsilon_{F}),
\end{equation}
with%
\begin{equation}
\frac{1}{\bar{v}_{\lambda}^{c}(\varepsilon_{F})}=\frac{1}{v_{\lambda}%
^{c}(\varepsilon_{F},\mathbf{R})}+\frac{1}{v_{\lambda}^{c}(\varepsilon
_{F},-\mathbf{R})}.
\end{equation}
Here, we have used $\mathbf{v}_{\lambda}^{c}(\varepsilon_{F},\pm
\mathbf{R)\parallel(\pm R)}$\ on the classical trajectories (cf. classical
trajectories in Sec. II.B). We further make the approximation $\digamma
_{jj^{\prime}}^{0}(\varepsilon,\mathbf{R})=\digamma_{jj^{\prime}}%
^{0}(\varepsilon_{F},\mathbf{R})$ for the slowly-varying function, then the
RKKY range function becomes%

\begin{equation}
J_{jj^{\prime}}(\varepsilon_{F},\mathbf{R})\approx\frac{J_{0}^{2}\bar
{v}_{\lambda}(\varepsilon_{F})}{2\pi R}\operatorname{Re}[\digamma_{jj^{\prime
}}^{0}(\varepsilon_{F},\mathbf{R})e^{i\phi_{s}^{c}(\mathbf{R})+i\phi_{-s}%
^{c}(-\mathbf{R})}]. \label{ARF}%
\end{equation}
The$\ $analytical formula Eq. \ref{ARF} implies that the RKKY interaction is
determined by the momentum states with group\ velocities parallel to the
connection vector $\mathbf{R}$ between two magnetic impurities and with the
energies equal to Fermi level $\varepsilon_{F}$ as shown by the straight solid
lines in Fig. \ref{RKKY}(b). In contrast, Eq. \ref{RF} has an explicit
integral over energy and an implicit integral over momentum for GF, so the
contributions of all momentum states below $\varepsilon_{F}$\ to the RKKY
interaction should be considered.\ Therefore, Eq. \ref{ARF} provides a
simplified version and a physically transparent expression of Eq.
\ref{RF}.\ Eq. \ref{ARF} has an immediate\ application, which leads to%
\begin{equation}
\mathbf{J}(\varepsilon_{F},\mathbf{R})=(\mathbf{J}(-\varepsilon_{F}%
,\mathbf{R}))^{\text{T}}.
\end{equation}
with the range function matrix
\begin{equation}
\mathbf{J=}\left[
\begin{array}
[c]{cc}%
J_{11} & J_{12}\\
J_{21} & J_{22}%
\end{array}
\right]
\end{equation}
by recalling Eq. \ref{EHS} (cf. symmetry discussions in Sec. II.B). This range
function relation originates from the inversion symmetry of velocity textures
for electrons and holes as shown by Fig. \ref{vcontour}, which favors our
discussions limited to electron doping case in our study.\ Eq. \ref{ARF}\ is
also used to plot the black solid lines in Figs. \ref{CRKKY}. At large
distances, the results of analytical formula and numerical calculations agree
very well with each other, so Eq. \ref{ARF} can capture the main features of
long-range RKKY interaction.

\begin{figure}[ptbh]
\includegraphics[width=\columnwidth,clip]{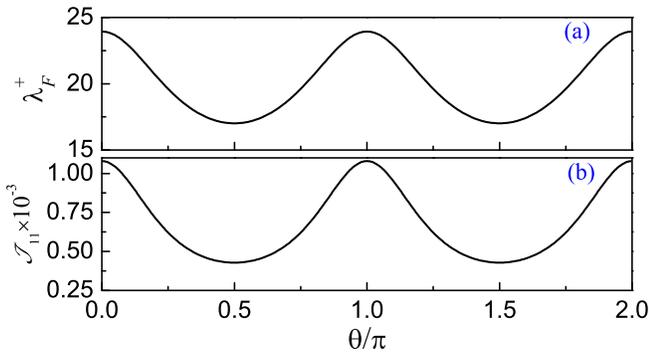}\caption{The oscillation
period $\lambda_{F}^{+}$ (a)\ and the envelop amplitude (b)\ of RKKY range
function. Here, $\varepsilon_{F}=0.1$.}%
\label{PA}%
\end{figure}

\subsection{Velocity-determined RKKY interaction and its implications}

Firstly, the decay rate of RKKY interaction is $J_{jj^{\prime}}(\varepsilon
_{F},\mathbf{R})\propto1/R^{2}$ since $\digamma_{jj^{\prime}}(\varepsilon
_{F},\mathbf{R})$ $\propto1/R$\ by referring to Eqs. \ref{MG2} and \ref{AGF},
this dimension-determined decay behavior is as that in conventional
two-dimensional electron gases \cite{FischerPRB1975,Beal-MonodPRB1987} and in
other doped Dirac materials with two-dimensional isotropic MDFs
\cite{SaremiPRB2007,BreyPRL2007,HwangPRL2008,SherafatiPRB2011a,KoganPRB2011,LiuPRL2009,ZhuPRL2011,AbaninPRL2011}%
.

Secondly, the oscillation period of $J_{jj^{\prime}}(\varepsilon
_{F},\mathbf{R})$\ is determined by the phase factor $\phi_{s}^{c}%
(\mathbf{R})+\phi_{-s}^{c}(-\mathbf{R})$, which is given by%

\begin{equation}
\lambda_{F}^{s}=\frac{2\pi R}{\left\vert \phi_{s}^{c}(\mathbf{R})+\phi
_{-s}^{c}(-\mathbf{R})\right\vert } \label{PED}%
\end{equation}
for the fixing Fermi level $\varepsilon_{F}$. To consider $\varepsilon
_{F}=0.1$,\ we use Eq. \ref{PED} to plot the oscillation period in Fig.
\ref{PA}(a) as the function azimuthal $\theta$ of the connection vector
$\mathbf{R}$. Fig. \ref{PA}(a) has the mirror symmetry about $y$-axis
originating from the symmetry of Hamiltonian \cite{PhysRevB.96.035410}, and
shows two peaks (valleys) along $\pm x$-axis ($\pm y$-axis).\ We defines the
peak-valley ratio as the anisotropy degree of the oscillation period.
Referring to Eq. \ref{AKC}, for $\varepsilon_{F}>0$, one can obtain
$\lambda_{F}^{+}(\theta=0)=\pi({\gamma_{1}^{2}-\gamma_{2}^{2}})^{1/2}%
/(\epsilon_{F}\gamma_{1})$ and $\lambda_{F}^{+}(\theta=\frac{\pi}{2}%
)=\pi(\gamma_{1}^{2}-\gamma_{2}^{2})/(\epsilon_{F}\gamma_{1})$, which both
show the inverse linear energy dependence of the oscillation period identical
to the case for the isotropic MDFs \cite{SherafatiPRB2011a,ZhuPRL2011}. In
particular, in Fig. \ref{PA}(a), $\lambda_{F}^{+}(\theta=0)=24$ and
$\ \lambda_{F}^{+}(\theta=\pi/2)=17$ when $\varepsilon_{F}=0.1$ for\ the RKKY
range function along $x$-axis and $y$-axis, respectively. As a result, the
anisotropy degree of the oscillation period is a constant $v_{x}/({v_{y}%
^{2}-v_{t}^{2}})^{1/2}\approx1.4$ independent of the Fermi energy
$\varepsilon_{F}$.

Thirdly, the envelope amplitude of the oscillating RKKY interaction is another
important feature beyond the decay rate and oscillation period. The envelope
amplitude depends linearly on Fermi level $\varepsilon_{F}$ by combing Eqs.
\ref{AGF} and \ref{ARF} and by using $\alpha_{s}\propto\varepsilon^{2}$ in Eq.
\ref{alphas}.\ In Fig. \ref{PA}(b), Eqs. \ref{AGF} and \ref{ARF} are used to
plot the envelope amplitude of RKKY interaction $J_{11}$\ as the function
azimuthal $\theta$ of $\mathbf{R}$. In analogy to the oscillation period, we
also define peak-valley ratio as the anisotropy degree for the envelope
amplitude, which can be extracted analytically from Eqs. \ref{AGF} and
\ref{ARF} and is a constant $v_{x}^{2}v_{y}^{2}/(v_{y}^{2}-v_{t}^{2}%
)^{2}\approx2.5$ independent of the Fermi energy $\varepsilon_{F}$.

As a result, the RKKY mediated by the anisotropic and tilted MDFs has
anisotropic features in its oscillation period and envelop amplitude but does
not change its decay rate. The spatial anisotropy is an usual feature of RKKY
interaction, at least on the atomic scale considering the discrete lattice
nature of host materials, and the spatially anisotropic RKKY interaction has
been reported in experiement\cite{ZhouNatPhys2010,PhysRevB.94.075137}.
Theoretically, there are quite a few studies on the close relation between
anisotropic RKKY interaction in long range (on the scale of Fermi wavelength)
and band features of various materials, e.g., III-V diluted magnetic
semiconductors\cite{PhysRevB.71.155206}, semiconductor quantum
wires\cite{ZhuPRB2010}, surface states on Pt(111)\cite{PatronePRB2012}, the
surface of the topological insulator Bi$_{2}$Se$_{3}$\cite{PhysRevB.85.054426}%
, spin-polarized graphene\cite{ParhizgarPRB2013}, three-dimensional Dirac
semimetals\cite{PhysRevB.93.094433}, three-dimensional electron gases with
linear spin-orbit coupling\cite{PhysRevB.96.115204}, and phosphorene
with\cite{1367-2630-19-10-103010} and without mechanical strain\cite{Zare2018}%
, and so on. However, comparing to the other materials, the anisotropic RKKY
interaction in 8-\textit{Pmmn} borophene is independent of the Fermi energy $\varepsilon_{F}$ but is velocity-determined, which has two
significant implications. On one hand, the velocity-determined RKKY
interaction fully determines the band dispersion described by three velocities
$v_{x,y,t}$ (cf. band dispersion of Eq. \ref{ED}), this\ provides an
alternative way to characterize the band dispersion. And the RKKY interaction
has also proposed to probe the topological phase transition in silicene
nanoribbon\cite{ZarePRB2016} and quasiflat edge modes in phosphorene
nanoribbon\cite{PhysRevB.98.205401,PhysRevB.97.235424}.\ On the other hand,
the anisotropy of RKKY interaction is expected to be tunable by choosing other
Dirac materials with different $v_{x,y,t}$ (e.g., quinoid-type graphene and
$\alpha$-(BEDT-TTF)$_{2}$I$_{3}$\cite{PhysRevB.78.045415}) and even by tuning
$v_{x,y,t}$ in the same Dirac material (e.g., graphene under
strain\cite{Naumis2017}). In Dirac materials, the RKKY interaction leads to
the ordering of magnetic impurities as discussed
theoretically\cite{LiuPRL2009,EfimkinPRB2014,PhysRevB.95.155414,PhysRevB.98.064425}
and demonstrated experimentally\cite{ncomms6349,ncomms12027,s41567-018-0149-1}%
. Based on the velocity-determined RKKY interaction mediated by anisotropic
tilted MDFs, new anisotropy-induced magnetic properties are expected and has
high tunability.\ Therefore, we hope this study is helpful to the physical
understanding of 8-\textit{Pmmn} borophene and its possible applications in
spin fields.

\section{Conclusions}

In this study, we investigate the RKKY interaction mediated by the anisotropic
and tilted MDFs in 8-\textit{Pmmn} borophene. In the long range, the RKKY
interaction is mainly contributed by the limited momentum states with group velocities
parallel to the connection vector $\mathbf{R}$ between two magnetic impurities
and with the energies equal to Fermi level $\varepsilon_{F}$. As a result, we
analytically demonstrate that the RKKY interaction in 8-\textit{Pmmn}
borophene has the usual decay rate as in the other two-dimensional materials,
but is anisotropic in its oscillation period and envelop amplitude with the
explicit velocity dependence $v_{x}/({v_{y}^{2}-v_{t}^{2}})^{1/2}$ and
$v_{x}^{2}v_{y}^{2}/(v_{y}^{2}-v_{t}^{2})^{2}$. The velocity-determined RKKY
interaction implies its usefulness to fully characterize the band dispersion
of 8-\textit{Pmmn} borophene and other similar Dirac materials, and its
tunability by engineering the anisotropic tilted MDFs. In addition, the
velocity-determined RKKY interaction should be observable in present experiment since
the evidence of RKKY interaction mediated by anisotropic MDFs on surface
Mn-doped Bi$_{2}$Te$_{3}$ has been reported through focusing interference
patterns observed by scanning tunneling microscopy\cite{PhysRevB.94.075137}.
This study is relevant to spatial propagation properties of novel anisotropic tilted MDFs, and shows the potential spin application of 8-\textit{Pmmn} borophene.

\textit{Note added}. Recently, we became aware of a related
preprint\cite{1811.08301}, which studies the RKKY interaction in
8-\textit{Pmmn} borophene along two specific directions.

\section*{Acknowledgements}

This work was supported by the National Key R$\&$D Program of China (Grant No.
2017YFA0303400), the NSFC (Grants No. 11504018, and No. 11774021), the MOST of
China (Grants No. 2014CB848700), and the NSFC program for ``Scientific
Research Center'' (Grant No. U1530401). We acknowledge the computational
support from the Beijing Computational Science Research Center (CSRC).


\end{document}